\documentclass[prl,twocolumn,showpacs,letterpaper,showpacs,superscriptaddress]{revtex4}
\usepackage{graphicx,amsmath,amssymb,amsfonts,latexsym,color,dcolumn,bm,epsfig,subfigure}

\begin{document}

\title{Reply to Comment [arXiv:0810.3244v1] by R.S. Decca et al. on ``Contribution of drifting carriers to the Casimir-Lifshitz and Casimir-Polder interactions with semiconductor materials".}

\author{Diego A. R. Dalvit}
\affiliation{Theoretical Division, Los Alamos National Laboratory, Los Alamos, NM 87545, USA}

\author{Steve K. Lamoreaux}
\affiliation{Yale University, Department of Physics,
P.O. Box 208120, New Haven, CT 06520-8120, USA}

\date{\today}

\begin{abstract}
We show that the claims expressed in the Comment \cite{Comment} against our paper 
\cite{ourPRL} are wrong and manifestly inconsistent with basic principles of statistical
physics.
\end{abstract}

\pacs{42.50.Ct, 12.20.-m, 78.20.-i}

\maketitle

The authors of the Comment \cite{Comment} write that  in our paper 
\cite{ourPRL} we stated that our approach is applicable only to intrinsic semiconductors. We made no such
a statement in our paper. Certainly we realize that the Boltzmann transport equation has been applied 
with great success throughout physics, and in \cite{ourPRL} we pointed out that our approach, using the 
{\it classical} Boltzmann equation, is limited to those situations where the electron gas is non-degenerate 
(Maxwell-Boltzmann statistics).  

The central point of the Comment \cite{Comment} is that supposedly our theory leads to thermodynamic and experimental inconsistencies.  It is argued that the alleged failure of our theory is due to ``the inclusion of irreversible diffusion processes violating thermal equilibrium into the standard Lifshitz theory which is derived under the condition of thermal equilibrium". This statement is obviously wrong and shows a clear misunderstanding of Lifshitz theory and quantum statistical physics. Indeed,  a dissipative component of the dielectric permittivity is required per Lifshitz formulation \cite{Lifshitz}, and the
fluctuations that lead to the Casimir force follow directly from the quantum fluctuation-dissipation theorem. The authors have confused irreversibility with
detailed balance: the fluctuating fields are dynamic by definition, and lead to (fluctuating) energy exchange between the material bodies and the electromagnetic
field; these fluctuations extend to absolute zero temperature, which of course cannot be attained. As explicitly mentioned in our Letter, the current driven by the fluctuating electric field is counterbalanced by diffusion. This compensation results in Einstein's relation between diffusion and mobility, and represents the {\it dynamic equilibrium} between a time-varying field and the charge distribution in the material.

A specific point of the Comment aimed at  demonstrating  the alleged thermodynamic inconsistencies of 
our theory deals with the extension of our technique to situations where the electrical 
conductivity of a material approaches zero at low temperatures, but the charge density in the material does not go to zero.
In the supporting calculations by some of the authors \cite{Galina2008} it is assumed that these charges remain ``free", and the 
decrease in conductivity is due to a decrease in mobility of these free charges. In this case, the so-called Casimir entropy $S$
does not approach zero as $T\rightarrow 0$, nor does $\partial S/\partial a$ approach zero in this limit ($a$ is the plate separation). So this part of the Comment on our paper can be interpreted as a discussion on how, in general, the charge carrier concentration and the conductivity of a material changes with temperature. 
This question is far beyond the scope of our paper; however, we do not agree with the statements in the Comment and in \cite{Galina2008} 
regarding this point. 
In particular, measurements in a restricted high-temperature interval \cite{Tomozawa1998}
have shown that for dielectrics with ionic conductivity {\it both} the conductivity {\it and} the charge carrier concentration decrease as temperature is lowered, the mobility remains nearly independent of temperature, and the conductivity activation
energy is primarily the energy needed to dissociate the ions. It is not clear to us
how the authors of the Comment extrapolate these experimental observations down to
$T=0$ and infer just the opposite behavior.

In the Comment it is argued that our treatment via the classical Boltzmann transport equation can be extended to degenerate systems 
(that should be described by the quantum Boltzmann equation and Fermi-Dirac statistics) by simply replacing the Debye-H\"uckel screening length in 
our equations in \cite{ourPRL} by the Thomas-Fermi screening length. Assuming this is the case, they claim that such an extension of our theory is 
incompatible with the Nernst theorem for perfect crystal lattices, as happens for the Drude 
model \cite{Bezerra2004}. In the absence of supporting arguments or calculations by the authors of the Comment,  we cannot give a definite opinion on such an extension and alleged inconsistencies. However, assuming they follow the same lines as  \cite{Bezerra2004}, we believe
that they are probably also wrong because \cite{Bezerra2004} incorrectly described the $T\rightarrow0$ behavior of a perfect crystal lattice  by the normal skin
theory for metals, instead of the appropriate anomalous skin theory.  As a general remark, we point out that the behavior of a model at low temperatures does not necessarily bear on its
validity or applicability at high temperatures.
Casimir systems are not unique in this regard: The entropy of an ideal gas, as described by the Sakur-Tetrode equation which takes into account quantum effects, diverges at zero temperature, yet this equation is known, by experiment, to be extremely accurate over a broad range.

As to the precision of various experimental results as pertains to the two figures in the Comment, in our opinion important systematic effects have not been properly taken care of in the electrostatic calibrations and
Casimir force residuals in the mentioned experiments
at the level of the claimed precision. 
The experiments reported in \cite{Cornell2007}
adequately addressed systematic effects and achieved an accuracy, in relation to the theory-experiment comparison, of about 10\%. 
This is  sufficient to verify the general validity of our theoretical approach;
related approaches have recently been published \cite{Pitaevskii2008,Svetovoy2008}.  
Although these experiments were a measurement of the Casimir-Polder force, the calculational techniques are similar to those of the Casimir force.  Our own experimental work using Ge plates, in which we uncovered a new systematic effect that has not been considered before in Casimir experiments, also indicates the general validity of our approach \cite{kim}.

\end{document}